\documentclass[iop]{emulateapj}

\slugcomment{To appear in ApJ}

\usepackage{graphicx}
\usepackage{natbib}
\shorttitle{Cold CO in the envelopes of FUors}
\shortauthors{K\'osp\'al et al.}

\begin{document}

\title{Cold CO gas in the envelopes of FU Orionis-type
    young eruptive stars}


\author{\'A. K\'osp\'al\altaffilmark{1,2},
  P. \'Abrah\'am\altaffilmark{1}, T. Csengeri\altaffilmark{3},
  Th. Henning\altaffilmark{2}, A. Mo\'or\altaffilmark{1},
  R. G\"usten\altaffilmark{3}}

\altaffiltext{1}{Konkoly Observatory, Research
  Centre for Astronomy and Earth Sciences, Hungarian Academy of
  Sciences, Konkoly-Thege Mikl\'os \'ut 15-17, 1121 Budapest, Hungary}

\altaffiltext{2}{Max-Planck-Institut f\"ur Astronomie, K\"onigstuhl
   17, 69117 Heidelberg, Germany}

\altaffiltext{3}{Max-Planck-Institut f\"ur Radioastronomie, Auf dem
  H\"ugel 69, 53121 Bonn, Germany}


\begin{abstract}
FUors are young stellar objects experiencing large optical outbursts
due to highly enhanced accretion from the circumstellar disk onto the
star. FUors are often surrounded by massive envelopes, which play a
significant role in the outburst mechanism. Conversely, the subsequent
eruptions might gradually clear up the obscuring envelope material and
drive the protostar on its way to become a disk-only T~Tauri
star. Here we present an APEX $^{12}$CO and $^{13}$CO survey of eight
southern and equatorial FUors. We measure the mass of the gaseous
material surrounding our targets. We locate the source of the CO
emission and derive physical parameters for the envelopes and
outflows, where detected. Our results support the evolutionary
scenario where FUors represent a transition phase from
envelope-surrounded protostars to classical T~Tauri stars.
\end{abstract}



\section{Introduction}

FU~Orionis-type objects (FUors) constitute a small group of young
stars characterized by large optical-infrared outbursts, attributed to
highly enhanced accretion \citep{hk96}. During these outbursts,
accretion rates from the circumstellar disk to the star are in the
order of 10$^{-4}$\,M$_{\odot}$/yr, three orders of magnitude higher
than in quiescence or in normal T~Tauri stars. FUors are natural
laboratories where not only enhanced accretion but enhanced mass loss
can be studied. Most FUors have optical jets, molecular outflows, and
optically visible ring-like structures on a 0.1\,pc scale that, in
some cases, might be connected to expanding shells thrown off during
previous outbursts \citep{mcmuldroch1993}.

Circumstellar envelopes are supposed to play a significant role in the
outbursts of FUors, partly by replenishing the disk material after
each outburst \citep{vorobyov2006}, partly by triggering the eruptions
\citep{bell1994}. For this reason, envelopes are not static, but
evolve with time. Based on the appearance of the 10$\,\mu$m silicate
feature, \citet{quanz2007} defined two categories of FUors: objects
showing the feature in absorption are younger, still embedded in a
circumstellar envelope; objects showing the silicate band in emission
are more evolved, with direct view on the surface layer of the
accretion disk. A similar evolutionary sequence was outlined by
\citet{green2006} based on the amount of far-infrared excess. These
studies suggest that FUors represent a fundamentally important
transition period during early star formation when the embedded
protostar clears away its enshrouding envelope to become a Class II
T\,Tauri star \citep{sw2001, green2013}.

Traditionally, a large part of our knowledge on FUor envelopes comes
from modeling broad-band spectral energy distributions of the dust
emission, based on spatially unresolved photometric data mainly at
infrared and submillimeter wavelengths. The gas component, however, is
typically much less studied. With the goal to obtain a general picture
of the molecular gas content, we perform a comprehensive and
homogeneous survey of all known FUors, by measuring millimeter CO
lines using single dish telescopes. In this paper we present
observations of the envelopes of eight southern and equatorial FUors,
and study the distribution and kinematics of the circumstellar gas,
including the characterization of the molecular outflows where
detected. Our data reveal the large variety and trends in the envelope
structures predicted by the evolutionary models.


\section{Observations}

\begin{deluxetable*}{ccccccccccc}[h!]
\tabletypesize{\scriptsize}
\tablecaption{CO observations of our targets.\label{tab:results}}
\tablewidth{0pt}
\tablehead{
\colhead{Name} & \colhead{Distance\tablenotemark{a}} & \colhead{$v_{\rm LSR}$}	& \colhead{F($^{12}$CO(3--2))} & \colhead{F($^{12}$CO(4--3))} & \colhead{F($^{13}$CO(3--2))} & \colhead{$\tau_{12}$} & \colhead{$\tau_{13}$} & \colhead{M$_{\rm tot}$} & \colhead{Outflow?} & \colhead{Si feature} \\
               & \colhead{(pc)}     & \colhead{(km\,s$^{-1}$)}  & \colhead{(Jy\,km\,s$^{-1}$)} & \colhead{(Jy\,km\,s$^{-1}$)} & \colhead{(Jy\,km\,s$^{-1}$)} &  		    &			   & \colhead{(M$_{\odot}$)} &  		  &		       }
\startdata
AR 6A/6B       & 800  & 5.3    & 3800$\pm$7   &  5190$\pm$39   &  1150$\pm$7     & 57  & 0.8 & 1.3\tablenotemark{b} & n & ?   \\
Bran 76	       & 1700 & 17.7   & 18.8$\pm$1   &  21.4$\pm$5.7  &  4.27$\pm$1.39  & 24  & 0.4 & 0.02		   & n & em  \\
HBC 494        & 460  & 4.3    & 3660$\pm$11  &  4780$\pm$50   &  1070$\pm$15    & 130 & 1.9 & 0.4		   & y & abs \\
HBC 687        & 400  & 17.2   & 173$\pm$6    &  164$\pm$22    &  37.0$\pm$6.7   & 15  & 0.2 & 0.01		   & n & em  \\
Haro 5a IRS    & 470  & 11.2   & 7940$\pm$9   &  13800$\pm$50  &  2990$\pm$8     & 76  & 1.1 & 1.2		   & y & abs \\
OO Ser	       & 311  & 8.1    & 15500$\pm$24 &  27800$\pm$140 &  3250$\pm$27    & 48  & 0.7 & 0.6		   & ? & abs \\
V346 Nor       & 700  & $-$3.0 & 2490$\pm$8   &  4780$\pm$33   &  383$\pm$8      & 52  & 0.8 & 0.3		   & y & abs \\
V900 Mon       & 1100 & 13.6   & 199$\pm$2    &  234$\pm$11    &  50.3$\pm$2.2   & 67  & 1.0 & 0.1\tablenotemark{b} & n & em
\enddata
\tablenotetext{a}{Distances are from \citet{audard2014} and
  \citet{reipurth1997b}.}
\tablenotetext{b}{For AR~6A/6B and V900~Mon, there is no distinct peak
  in the CO emission at the stellar position, therefore the CO
  emission (and masses given here) are most likely not associated with
  these sources.}
\end{deluxetable*}

Table~\ref{tab:results} lists the targets selected for our study from
the list of \citet{audard2014}. We used the FLASH$^+$ receiver
\citep{klein2014} at the APEX telescope {\citep{gusten2006}} to
measure the $^{12}$CO(3--2), $^{13}$CO(3--2), and $^{12}$CO(4--3)
lines towards our targets between 2014 August 23 -- 28. APEX is a
12\,m diameter millimeter-wave telescope located on the Llano de
Chajnantor at 5104\,m altitude in the Chilean Atacama
desert. FLASH$^+$ is a dual-frequency heterodyne receiver, operating
simultaneously in the 345\,GHz and the 460\,GHz atmospheric windows,
providing 4\,GHz bandwidth in each sideband. The lower frequency
channel was tuned to 344.2\,GHz in USB to cover the $^{13}$CO(3--2) at
330.588\,GHz, and the $^{12}$CO(3--2) at 345.796\,GHz,
respectively. The higher frequency channel was tuned to the
$^{12}$CO(4--3) line at 461.041\,GHz in USB. We used the XFFTS
backends providing a nominal 38\,kHz spectral resolution for the 3--2
lines and 76\,kHz for the 4--3 line. For each target,
90$''{\times}$90$''$ on-the-fly (OTF) maps were obtained, using a
relative reference off position 1000$''$ away in right ascension. We
started each observation by checking in total power mode whether the
off positions are clean. If needed, we modified the OFF position
(1200$''$ for V900~Mon and 800$''$ for Bran~76), until we made sure
that there is no CO emission at the velocity of the target.

A first order baseline was removed from the spectra. The data were
calibrated using a main beam efficiency of 0.73 and 0.60 at 352 and
464\,GHz, respectively, and the values were converted to Jy using
41\,Jy\,K$^{-1}$ and 48 \,Jy\,K$^{-1}$ at 352 and 464\,GHz,
respectively. The rms noise level calculated for the line-free
channels in 1\,km\,s$^{-1}$ bins is 0.8\,Jy for the $^{12}$CO(3--2)
and $^{13}$CO(3--2) lines, and 2.3\,Jy for the $^{12}$CO(4--3)
line. The telescope's beam is $19$\farcs$2$, and $15$\farcs$3$ at the
corresponding frequencies.


\section{Results and analysis}

\begin{figure*}
\includegraphics[angle=90,scale=0.9]{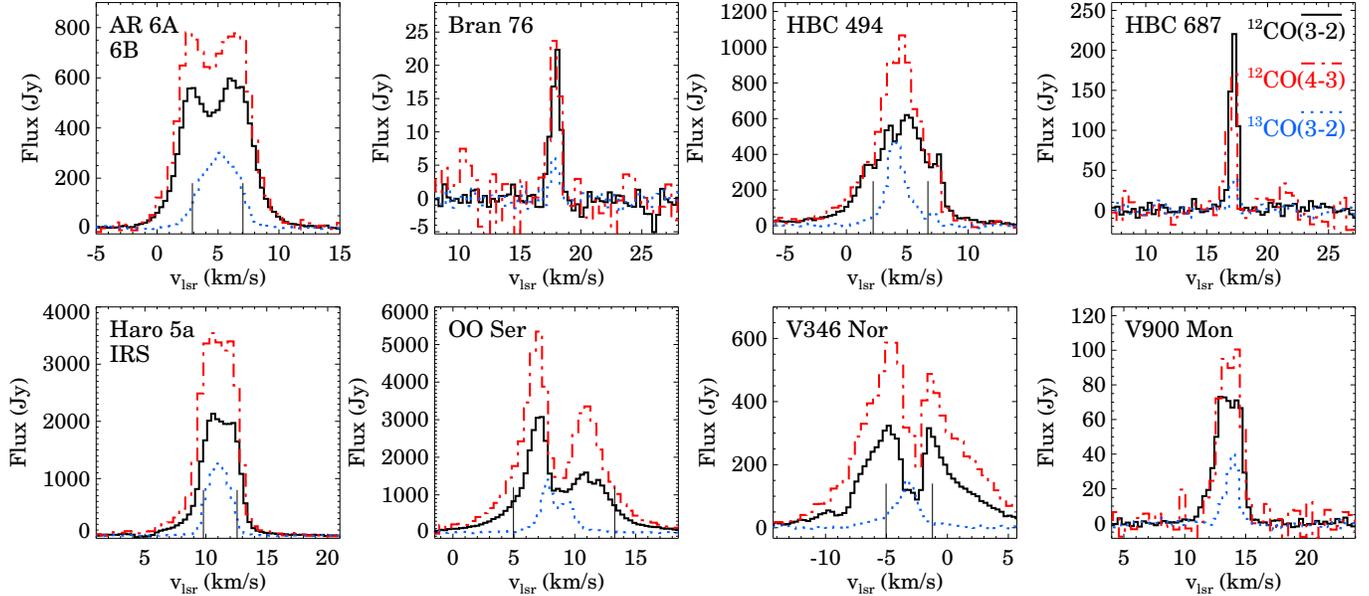}
\caption{CO line profiles of our targets observed with APEX.  Fluxes
  were integrated within a 10\,000\,au radius centered on the nominal
  position of our targets. The line wings marked by the vertical lines
  indicate possible outflows.
\label{fig:fuorco}}
\end{figure*}

\begin{figure*}
\includegraphics[angle=90,scale=0.915]{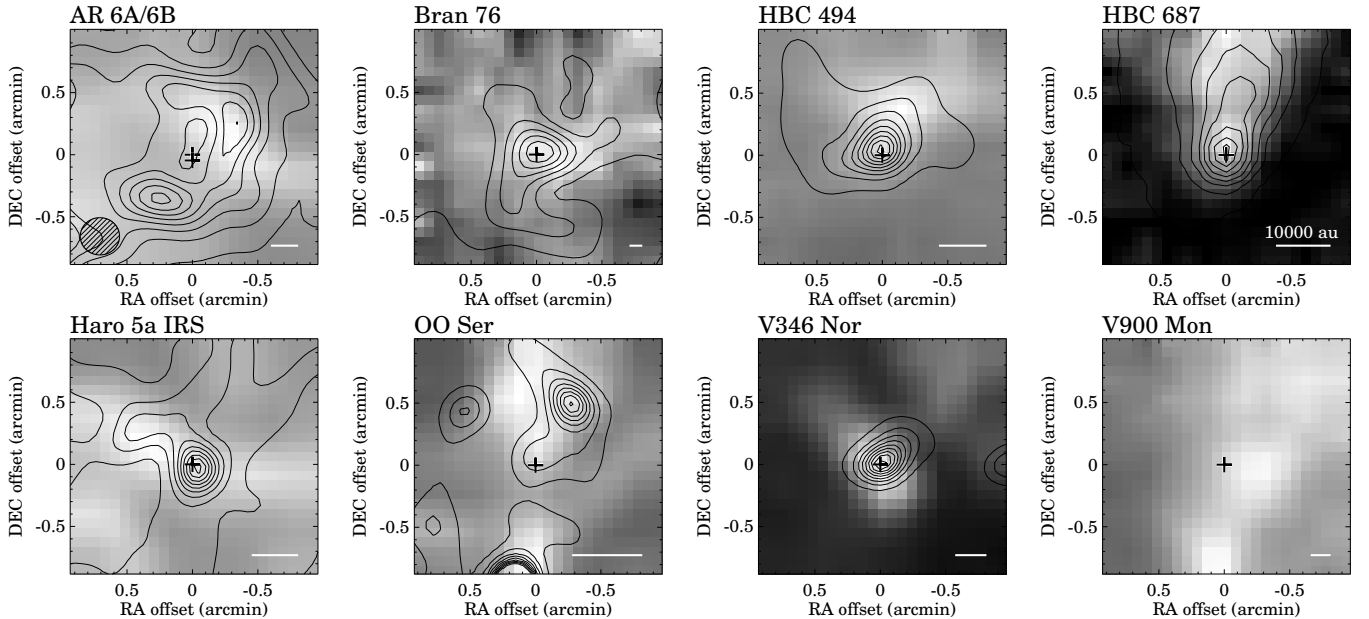}
\caption{Integrated CO intensity maps of our targets for the
  $^{12}$CO(3--2) line observed with APEX (grayscale) and 250$\,\mu$m
  continuum emission from Herschel (contours). Plus signs mark the
  stellar position, while the hatched circle shows the Herschel beam
  size.
\label{fig:mom0}}
\end{figure*}

CO emission for all targeted isotopologues and transitions were
detected in our maps. In Fig.~\ref{fig:fuorco} we show the CO line
profiles integrated within a 10\,000\,au radius centered on the
nominal position of our targets, while Fig.~\ref{fig:mom0} shows the
total CO line intensity maps integrated for the velocity channels
where at least a 3$\sigma$ signal was detected. The flux-weighted
average $v_{\rm LSR}$ velocities for the CO emission are listed in
Table~\ref{tab:results}. The velocity-integrated line fluxes for same
spatial areas (within 10\,000\,au) are also given in
Table~\ref{tab:results}. We used the optically thin $^{13}$CO lines to
convert the observed line fluxes to total gas masses assuming local
thermodynamic equilibrium, using 20\,K temperature, a
$^{13}$CO/$^{12}$CO abundance ratio of 69 \citep{wilson1999} and
$^{12}$CO/H$_2$ abundance ratio of 10$^{-4}$ \citep{bolatto2013}. We
note that if we use 50\,K instead of 20\,K, the masses would be a
factor of 1.06 lower, and if we used 15\,K instead of 20\,K, the
masses would be a factor of 1.29 times higher.

\paragraph{Line profiles} Figure~\ref{fig:fuorco} demonstrates that
out of our sample, Bran~76 and HBC~687 show the narrowest lines, the
FWHM is only about 0.7--0.8\,km\,s$^{-1}$. V900~Mon is somewhat
broader, while the rest of the targets show very broad lines, and
prominent line wings in the $^{12}$CO lines. For most of the sources
(AR~6A/6B, HBC~494, Haro~5a~IRS, V346~Nor, and V900~Mon), the
$^{13}$CO line is single-peaked, while the $^{12}$CO lines are either
flat-topped or show self-absorption. This suggests that $^{12}$CO is
optically thick. The same is true for the two targets with the narrow
lines, where the ratio of the $^{12}$CO(3--2) to the $^{13}$CO(3--2)
line peaks suggest a maximum optical depth of $\tau_{12}$=15--24 for
the former and $\tau_{13}$=0.2--0.4 for the latter. For the rest of
the targets, the line peaks indicate somewhat larger optical depths,
in the 50--130 range for $\tau_{12}$ and in the 0.7--1.9 range for
$\tau_{13}$. The line profile of OO~Ser is different from the other
sources, because even the $^{13}$CO line seems to be
double-peaked. Because the line ratios do not indicate extraordinarily
high optical depths, we suspect that in this case several different
velocity components are superimposed along the line of sight. The
observed diversity of the line profiles seems to be a characteristic
of the FUor class. \citet{evans1994} presented single-dish CO line
data for a sample containing both northern and southern FUors. In
RNO~1B, V1735~Cyg, and V346~Nor they detected self-absorbed $^{12}$CO,
while Z~CMa, V1057~Cyg, and V1515~Cyg displayed narrow, single-peaked
$^{12}$CO emission. The $^{13}$CO line was single-peaked in all of
their sources, similarly to our results.

\paragraph{Integrated emission maps} For all of our targets there is
some CO emission towards the stellar position, but there is also
significant confusion from extended emission. Bran~76, HBC~494,
HBC~687, Haro~5a~IRS, OO~Ser, and V346~Nor, where the CO emission
peaks at the stellar position, are clearly detected. For AR~6A/6B, and
V900~Mon, the CO emission peak is offset, so it cannot be
unambiguously associated with the star. In any case, the masses we
calculated should be considered as upper limits for the envelope
masses due to confusion. Three of our sources were targeted by
\citet{sw2001} in 850$\,\mu$m and 1.3\,mm continuum. While Bran~76 was
undetected, for HBC~494 and V346~Nor they give envelope masses
assuming 50\,K for the dust temperature. Their values
(0.1\,M$_{\odot}$ for HBC~494 and 0.5\,M$_{\odot}$ for V346~Nor) are
in good agreement with our mass estimates from the CO line fluxes
using 50\,K (0.4\,M$_{\odot}$ for HBC~494 and 0.3\,M$_{\odot}$ for
V346~Nor).

\paragraph{Comparison with dust continuum emission} In
Fig.~\ref{fig:fuorco} we overplotted with contours the 250$\,\mu$m
emission using Herschel/SPIRE data from the Herschel Science Archive
(proposal IDs: KPGT\_fmotte\_1, KPGT\_pandre\_1,
OT1\_maudar01\_1). Herschel at this wavelength had a similar beam size
(18$''$) to our APEX beam for the J=3--2 lines
(19$\farcs$2). Generally there is a good agreement between the
continuum and the CO line maps, although the continuum peaks are more
prominent than the CO peaks, and there is less extended emission in
continuum than in CO. Just like in CO, Bran~76, HBC~494, HBC~687,
Haro~5a~IRS, and V346~Nor are clearly detected in continuum, OO~Ser is
marginally detected, while AR~6A/6B seem to be sitting in the middle
of a cavity. Unfortunately, V900~Mon was not observed by Herschel.

\begin{figure*}
\includegraphics[angle=90,scale=0.915]{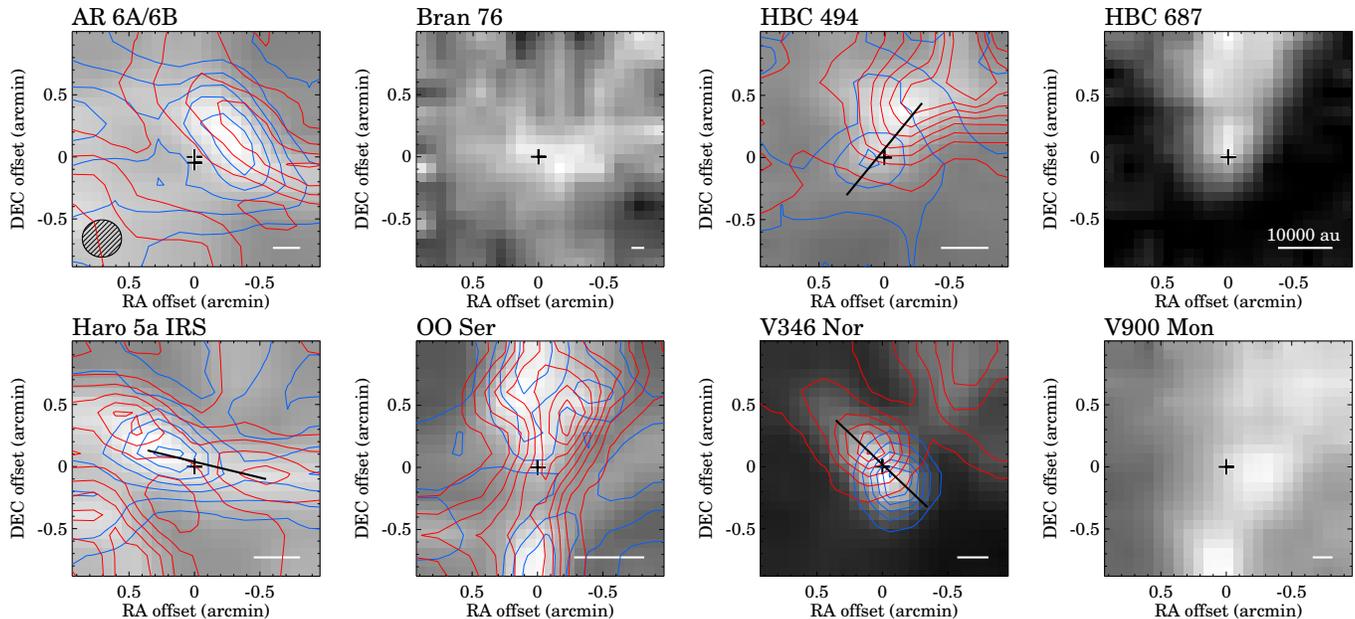}
\caption{Integrated CO intensity maps of our targets for the
  $^{12}$CO(3--2) line observed with APEX (grayscale). The red and
  blue contours show redshifted and blueshifted emission integrated in
  the velocity ranges indicated in Fig.~\ref{fig:fuorco} (contour
  levels are at 3, 6, 9, \dots $\sigma$). Thick black lines mark the
  directions of the detected bipolar outflows. Plus signs mark the
  stellar position, while the hatched circle shows the Herschel beam
  size.
\label{fig:redblue}}
\end{figure*}

\paragraph{Outflows} Some $^{12}$CO lines have
high-velocity wings, indicated by the vertical lines in
Fig.~\ref{fig:fuorco}. We integrated the emission for the red-shifted
and blue-shifted parts and plotted the resulting maps with red and
blue contours in Fig.~\ref{fig:redblue}. We detected clear signs for
outflows in HBC~494, Haro~5a~IRS, and V346~Nor. OO~Ser may also drive
an outflow, but at this spatial resolution, the detection is only
tentative due to confusion in the area. For the three unambiguously
detected outflows, we measured the masses, momenta, and energies of
the blue and red lobes, following the method and equations presented
in \citet{dunham2014}. The results, listed in Tab.~\ref{tab:outflow},
are calculated both in the optically thin approximation, and by
correcting for the optically thick emission using the
$(1-e^{-\tau_{12}})/\tau_{12}$ correction factor, where $\tau_{12}$
was calculated from the $^{12}$CO / $^{13}$CO line ratio in each
velocity channel. The outflow masses, momenta, and energies of the
FUors fall into the upper 30\% compared to the distribution of these
values measured by \citet{dunham2014} for a sample of 28 outflows
driven by low-mass protostars. The outflow of V346~Nor was already
detected by \citet{evans1994} in $^{12}$CO(3--2) and by
\citet{reipurth1997} in the $^{12}$CO(1--0), revealing a similar
morphology of the outflowing gas as our data. \citet{lee2002} observed
HBC~494 in $^{12}$CO(1--0). Their channel maps at $\approx$15$''$
resolution look very similar to ours. Haro~5a~IRS and its surroundings
were observed in $^{12}$CO(1--0) and $^{12}$CO(3--2) by
\citet{takahashi2006} and \citet{takahashi2008}. They clearly detected
the CO outflow from Haro~5a~IRS, and found an embedded protostellar
candidate, MMS 7-NE, which also drives an outflow. This complex
spatial and velocity structure of the CO emission is also reflected in
our observations.


\section{Discussion and conclusions}

Figure 1 and Table 1 reveal a striking variety of envelope properties
within our sample. One of these properties is envelope
mass. Haro~5a~IRS, OO~Ser, V346~Nor, and HBC~494 contain a significant
amount of gas ($>$0.3\,M$_{\odot}$). HBC~678 and Bran~76 are
associated with only 0.01-0.02\,M$_{\odot}$ of material. Having no
associated CO peaks, AR~6A/6B and V900~Mon probably also have low-mass
envelopes. Interestingly, other envelope parameters suggest an almost
identical division of the sample. Envelopes with higher mass exhibit
broader lines, while the low-mass envelopes have narrower lines or
remain undetected towards the source. Temperatures calculated from the
ratio of the $^{12}$CO(4--3) and $^{12}$CO(3--2) line ratios show that
the low-mass envelopes are typically cold (5-7\,K), while the
higher-mass envelopes are warm ($>$40\,K). Outflows are only detected
from sources with higher-mass envelopes. By observing a large sample
of low-mass protostars, \citet{jorgensen2009} found that the envelope
mass decreases sharply from typically 1\,$M_{\odot}$ in Class\,0
objects to ${<}\,0.1\,M_{\odot}$ in the Class\,I phase. Placing our
targets into this evolutionary scheme, FUors with higher envelope
masses represent a very early evolutionary phase. The FUors with low
envelope masses or upper limits may be close to the end of the
Class\,I phase.

As we summarized in the Introducion, \citet{quanz2007} proposed a
different way to order FUors into an evolutionary sequence: objects
showing silicate absorption at 10$\,\mu$m are younger, still embedded
in their opaque envelopes, while those showing silicate emission are
more evolved. We checked the Cornell Atlas of Spitzer/IRS
Sources\footnote{http://cassis.sirtf.com/} for mid-infrared spectra,
and found that HBH~494, Haro~5a~IRS, OO~Ser, and V346~Nor have
silicate absorption feature, while Bran~76, HBC~687, and V900~Mon
exhibit emission (see also Tab.~\ref{tab:results}). Our division based
on the CO gas properties of our targets correlates well with the
division based on the 10$\,\mu$m silicate feature. Objects possessing
massive gas envelopes exhibit silicate absorption, while those with
lower-mass envelopes show silicate emission
(Tab.~\ref{tab:results}). This conclusion suggests a parallel
evolution of the circumstellar dust and gas in FUors.

When young stellar objects transition from the embedded to the
disk-only phase, they clear away their envelopes to become an
optically visible star. According to the hypothesis of
\citet{quanz2007}, repetitive FUor outbursts may drive this process,
since a thick obscuring envelope produces an absorption feature, while
for an emission feature, large opening angle for the polar cavity in
the envelope is needed to provide a clear line-of-sight to the inner
disk (see also \citealt{kenyon1991}, \citealt{green2006}). FUor
outbursts gradually widen the outflow cavity due to the enhanced
outflow activity during the eruptions.

During a typical FUor outburst 0.01$\,M_{\odot}$ mass is accreted onto
the central star \citep{hartmann2008}. Depending on the length of the
quiescent periods, a similar amount of material may be accreted
between the outbursts. This is comparable to the envelope mass found
in our more evolved subsample. The mass reservoir in these systems to
replenish the disk after an outburst is very small. Therefore, these
objects are probably very close to the transition to the disk-only
phase and may represent the links between the Class\,I and Class\,II
phases of protostellar evolution.

\begin{table}
\begin{center}
  \caption{Outflow masses (M), momenta (P) and energies (E).
    \label{tab:outflow}}
\begin{tabular}{lccc}
\tableline\tableline
Parameter                                  & HBC 494              & Haro 5a IRS          &  V346 Nor            \\
\tableline
\multicolumn{4}{c}{optically thin}\\
\tableline
M (blue)   $M_{\odot}$ 	           & 0.004		  & 0.023		 & 0.020		\\
M (red)    $M_{\odot}$ 	           & 0.007		  & 0.020		 & 0.054		\\
P (blue)   $M_{\odot}$\,km\,s$^{-1}$ & 0.016		  & 0.039		 & 0.076  		\\
P (red)    $M_{\odot}$\,km\,s$^{-1}$ & 0.017		  & 0.042		 & 0.228  		\\
E (blue)   erg 		           & 7.8$\times$10$^{41}$ & 8.2$\times$10$^{41}$ & 3.8$\times$10$^{42}$ \\
E (red)    erg 		           & 1.2$\times$10$^{42}$ & 1.0$\times$10$^{42}$ & 1.3$\times$10$^{43}$ \\
\tableline
\multicolumn{4}{c}{optically thick}\\
\tableline
M (blue)  $M_{\odot}$ 	           & 0.021		  & 0.311		 & 0.100		\\
M (red)   $M_{\odot}$ 	           & 0.053		  & 0.131		 & 0.092		\\
P (blue)  $M_{\odot}$\,km\,s$^{-1}$ & 0.061		  & 0.417		 & 0.281  		\\
P (red)   $M_{\odot}$\,km\,s$^{-1}$ & 0.166		  & 0.244		 & 0.312  		\\
E (blue)  erg 		           & 2.1$\times$10$^{42}$ & 6.2$\times$10$^{42}$ & 9.3$\times$10$^{42}$ \\
E (red)   erg 		           & 5.6$\times$10$^{42}$ & 4.8$\times$10$^{42}$ & 1.5$\times$10$^{43}$ \\
\tableline
\end{tabular}
\end{center}
\end{table}


\acknowledgments

This work was supported by the Momentum grant of the MTA CSFK
Lend\"ulet Disk Research Group, and the Hungarian Research Fund OTKA
grant K101393. T.Cs. acknowledges
support from the \emph{Deut\-sche For\-schungs\-ge\-mein\-schaft,
  DFG\/} via the SPP (priority programme) 1573 `Physics of the ISM'.

{\it Facilities:} \facility{APEX}.


\end{document}